# The Case for Contextual Copyleft: Licensing Open Source Training Data and Generative AI

Grant Shanklin[1,2], Emmie Hine[1,3,4], Claudio Novelli[1], Tyler Schroder[1,2], Luciano Floridi[1,3]

[1] Digital Ethics Center, Yale University, 85 Trumbull St, New Haven, CT 06511

[2] Department of Computer Science, Yale University, 51 Prospect St, New Haven, CT 06511

[3] Department of Legal Studies, University of Bologna, Via Zamboni 27/29, 40126 Bologna, Italy

[4] Center for IT & IP Law, KU Leuven, Sint-Michielsstraat, 6 30000 Leuven, BE

Email for correspondence: grant.shanklin@yale.edu

**Abstract**

The proliferation of generative AI systems has created new challenges for the Free and Open Source Software (FOSS) community, particularly regarding how traditional copyleft principles should apply when open source code is used to train AI models. This article introduces the Contextual Copyleft AI (CCAI) license, a novel licensing mechanism that extends copyleft requirements from training data to the resulting generative AI models. The CCAI license offers significant advantages, including enhanced developer control, incentivization of open source AI development, and mitigation of open-washing practices. This is demonstrated through a structured three-part evaluation framework that examines (1) legal feasibility under current copyright law, (2) policy justification comparing traditional software and AI contexts, and (3) synthesis of cross-contextual benefits and risks. However, the increased risk profile of open source AI, particularly the potential for direct misuse, necessitates complementary regulatory approaches to achieve an appropriate risk-benefit balance. The paper concludes that when implemented within a robust regulatory environment focused on responsible AI usage, the CCAI license provides a viable mechanism for preserving and adapting core FOSS principles to the evolving landscape of generative AI development.

## 1. Introduction

With the recent increase in the number (Gozalo-Brizuela and Garrido-Merchan 2023) and power (Bandi, Adapa, and Kuchi 2023) of generative AI systems, the open source community faces new challenges in responding effectively. There is considerable discussion regarding the adoption of open source generative AI models, particularly given the substantial opportunities and risks that accompany making these systems open (Eiras, Petrov, Vidgen, Schroeder, et al. 2024; Floridi et al. 2025). Some participants in the debate suggest shifting the focus away from open source generative AI, emphasizing the need for responsible AI through behavioral use licensing instead (Contractor et al. 2022). Questions persist about how the open source community should manage models trained on open source code (a practice notably illustrated by GitHub's CoPilot, as discussed in Section 4.2). This dialogue focuses on preserving and adapting the rules, principles, and practices of the Free and Open Source Software (FOSS) movement to navigate the evolving technical landscape introduced by generative AI.

One way the FOSS movement has historically protected free software is through "copyleft" licenses. These licenses require that derivative works be distributed under the same open source license as the original work. This safeguards free and open source software because copyleft propagates licenses from source material to derivative works, creating a viral effect where both original and derivative works remain free and open (Phillips 2007). If attributes such as "freedom" and "openness" are desirable for generative AI systems, then copyleft may provide a promising mechanism for the FOSS community to uphold traditional software freedoms and extend their normative commitments into the emerging field of generative AI.

This paper examines whether such an extension of copyleft principles is desirable and explores potential mechanisms for achieving it. The *Contextual Copyleft AI (CCAI) license* is proposed, which may cover both traditional software and generative AI contexts. Then, the challenges of evaluating licenses that cross contextual boundaries are discussed, considering the legal and policy considerations in each context. This leads to an argument about how such a license should be evaluated. Next, this structured evaluation is completed for the CCAI license. The analysis concludes that the license could be beneficial, particularly when it is used within a strong regulatory environment. The license provides developers with greater control over how their code is used, encourages the development of open source generative AI systems, and addresses concerns related to openwashing. Regulating the responsible use of AI would address the risk-benefit imbalance inherent in copyleft, as highlighted by the evaluation. This license highlights emerging cases where carefully constructed and thoughtfully evaluated copyleft licenses cross contextual boundaries, offering essential insights into how to preserve and adapt free and open source norms, principles, and practices in a changing technological landscape.

**2. The Contextual Copyleft AI (CCAI) License**

This section proposes a new open source license that harnesses the features of copyleft to enable developers to allow only open source generative AI models to train on their code. The copyleft scheme enables copyright holders to permit users of their software to use, modify, and redistribute their code in any way, provided that the modified or redistributed version is licensed under the same terms as the original work (Stallman 2002). Fundamentally, copyleft is a mechanism that leverages the existing copyright system to protect and promote the freedom of users to run, copy, distribute, study, change, and improve software.

We will refer to the proposed license as the Contextual Copyleft AI (CCAI) license, a name that draws on this license's ability to cross technical and legal contexts. This ability necessitates a nuanced evaluation of this license in the sections that follow. The full license text and a clause that can be added to existing open source licenses are included in the appendix of this paper. The CCAI license consists of three main components:

1. Software freedom: Users can run the software, modify the code to fit their needs, and distribute copies of the software.
2. Copyleft requirement: Any verbatim copy, modified version, derivative work, or model trained on this code must also be wholly licensed under CCAI with no additional restrictions. This includes a description of the training data, the code used to train and run the model, and the model's parameters for generative AI models.[1]
3. Distribution clarification: Any distributed CCAI software must include source code or provide a way to access it. Providing access to software over the network is considered equivalent to distributing the software.

The CCAI license incorporates a copyleft, share-alike clause to propagate open source requirements from training data to the AI model itself. It adheres to the open source norm of imposing terms on distribution without using code to protect software freedom via an extended definition of "distribute." Furthermore, it enhances existing open source copyleft licenses by explicitly defining terms related to generative AI, such as including non-code aspects of an AI model in the definition of "entire work," which helps enable developers to freely share their code online while preventing the unconstrained training of AI models with their work. This is based on the premise that using code to train generative AI models is not considered fair use, as discussed above. Note that adding the CCAI clause to an existing open-source license effectively creates a new variant of the license. Care should be taken to ensure compatibility with other open-source licenses, as some existing licenses (such as the GPL) do not permit additional restrictions beyond their stated terms.

---

[1] The description of the training data is inspired by the Open Source Initiative's "Open Source AI" definition found at https://opensource.org/ai/open-source-ai-definition. Description should generally include what data was used to train the model, where the data originated, how it was selected and obtained, and discussion of any training or fliting methods.

The first component of the license is software freedom, defined as allowing users to "run the software, modify the code to fit their needs, and distribute copies of the software." This establishes the license's goal and ensures that the usage rights of the code are granted to the software user unconditionally, in accordance with all other open source licenses. This provision is essential for making the license open source and supporting all four software freedoms.

The second component of the license is the copyleft requirement, which mandates that "any verbatim copy, modified version, derivative work, or model trained on this code must also be wholly licensed under CCAI, with no additional restrictions." This requirement closely resembles the copyleft stipulations in the GNU General Public License (GPL) and Affero General Public License (AGPL) licenses,[2] with the added categorization of AI models as part of derivative works. This definition would impose share-alike terms on AI models if using code for training is not deemed fair use (if such training were fair use, the original copyright holder's license conditions could not apply to the model). However, while the courts have yet to determine whether this training constitutes fair use, it is crucial to include this clarification. The most significant addition to the fundamentals of the GPL and AGPL is the second part of this component, which outlines the specific elements that must be licensed under CCAI, particularly for AI models. The open sourcing must encompass "a description of the training data, the code used to train and run the model, and the model's parameters" under the same license without additional restrictions. This definition was based on the Open Source Initiative's definition of Open Source AI.[3] This definition helps combat open-washing, because it specifies the minimum threshold for what counts as an open source model in the eyes of a copyleft license.

The third component of the license is the distribution clarification, a clause taken from the AGPL license that categorizes accessing software over the network as distributing the software being accessed. This clarification is essential to ensure that model users' access via the internet is considered

---

[2] These are two popular OSS licenses, with AGPL targeted towards network-facing software.
[3] https://opensource.org/ai/open-source-ai-definition

"distributed" rather than being used by the developer. Otherwise, a model trained on data licensed under CCAI could allow users to access it through a website without requiring the model to be open source, as these obligations come into effect upon distribution.

**3. Evaluating Licenses Propagating Between Contexts**

Licenses that are designed to propagate between works in different technical and legal contexts pose unique challenges in evaluating their legal and policy implications. This section argues that the changing context of such licenses necessitates three components. First, a legal analysis is conducted to determine whether the propagation mechanism is permissible within the existing legal framework. Second, a policy analysis is performed to assess the individual benefits of copyleft within each context. This analysis must then be synthesized to determine the overall desirability of such licenses. While the necessity of the legal analysis is straightforward, the policy consideration is more nuanced. An open source license may be desirable in its original context and be able to propagate to another context, but the license may not be desirable in that new context. There are additional considerations, as the primary justification for copyleft licenses is necessarily context-dependent; therefore, the rationale for a license will change as it propagates between contexts. After introducing this framework for evaluating licenses that propagate across different technical and legal contexts, it will be applied to the CCAI license proposed above.

Technical contexts refer to sets of software that encompass work with similar risks of being made open source. For instance, traditional software refers to software that is commonly open sourced today, including desktop applications, computer science tools, and server software. The risks associated with traditional open source software may include security vulnerabilities, dependency management challenges, and concerns regarding project sustainability (Zajdel, Costa, and Mili, 2022). Another context could be generative AI because the risks of releasing a generative AI model as open source would be fundamentally different; the risks are focused on the harm that could be caused either by humans using the model or by a misaligned model itself (Amodei et al. 2016). These technical contexts may or may not exist within the same legal contexts. Typically, technologies operating in various technical contexts are governed by the same overarching legal framework, although specific regulatory requirements may differ.

The first component in evaluating a license designed to propagate between works in different contexts is a legal evaluation of the feasibility of the propagation mechanism. While the necessity of such analysis is self-evident, the examination itself may prove complex. This evaluation must primarily establish two elements: first, that the work to which the license propagates qualifies as a derivative of the original work under applicable copyright law; and second, that the incorporation of the original work within the derivative work falls outside the scope of the fair use doctrine. If these conditions are satisfied, it can generally be concluded that the license governing the original work may impose legal obligations on its use within the derivative work, including the enforcement of share-alike provisions.

The second component is a policy evaluation of the use of copyleft in each context individually. This must be done in each context individually because different technical contexts present fundamentally different risk profiles that affect the cost-benefit analysis underlying the justification of free software. These varying risks mean that the same copyleft license may be justified in one context but unjustified in another, even when applied to functionally related software. The conclusions from the separate policy evaluations must be considered jointly in determining whether such a license should be used.

Copyleft licenses are politically justified by their ability to protect and promote free and open source software.[4] Thus, the justification for using copyleft within a specific context depends on the rationale for using free and open source software in that context. If the use of free software generally can't be shown to be beneficial in a technical and legal context, then copyleft can't be shown to be beneficial either. Furthermore, the justification of free software must also consider the risks associated with making a piece of software open source. Thus, when technical contexts change, the specifics of the justification for free software, and therefore the justification for using copyleft, must change as well. The overarching argument in support of free software may remain unchanged; however, the specifics of the argument must evolve either way. For instance, copyleft on a traditional piece of software may be justified by a pragmatic argument regarding the quality of software produced by free and open source communities. If this license propagates to a generative AI system, the justification may or may not change, remaining pragmatic or taking on moral terms instead. However, the specifics of the argument *must* change due to the different fundamental risks associated with making the software open source. Therefore, the policy evaluation must be conducted individually for each technical context.

The third component is a policy analysis of copyleft propagation as a whole. While the policy analysis justifying the use of copyleft and open source must be conducted individually in each context, the results should be considered collectively to make a final determination as to whether the license should be used or not. This synthesis must evaluate the specific benefits and drawbacks of the propagation mechanism itself, including: whether it empowers developers by giving them appropriate control over how their work is used across contexts; whether it creates positive incentives for open source development in the new context; whether it helps maintain the integrity of open source principles as they extend to new technologies; and whether these benefits of cross-context propagation outweigh any tensions created by applying the same license across different risk profiles. The synthesis should conclude with a holistic assessment of whether the propagation serves the overarching goals of the FOSS community.

---

[4] https://www.gnu.org/licenses/licenses.html

**4. Contextual Copyleft from Traditional Software to Generative AI**

In this section, we evaluate the proposed CCIA license using the three-part analysis described in the previous section. First, we examine the legal mechanisms through which copyleft licenses could extend from training data to AI models, focusing on the critical question of whether AI training constitutes fair use under US copyright law. This legal analysis proceeds under the assumption that such training does not automatically qualify as fair use, allowing us to explore how licensing terms might apply.

Second, we evaluate the use of copyleft licenses in each context individually—traditional software and generative AI—to understand how the justification for these licenses changes across contexts. This comparative analysis reveals significant differences in the risk profiles between traditional open source software and open source AI, particularly regarding the potential for direct misuse and harm generation.

Finally, we synthesize these analyses to demonstrate that a copyleft license that propagates to AI systems offers three key benefits: it grants developers greater agency over how their code is used in AI development, stimulates the creation of open source generative AI systems, and combats "open-washing" practices. Practical enforcement of the CCAI license against AI models trained on publicly available data might be challenging due to difficulties in detecting unauthorized usage. However, even in cases where enforcement is challenging, the existence of clearly articulated license terms can serve as an effective deterrent against non-compliant use by law-abiding entities. We also acknowledge that the increased risks associated with open source AI, particularly the potential for direct misuse, require consideration of complementary regulatory approaches. We conclude by arguing that responsible AI regulation can work alongside copyleft licensing to better align the risks and benefits of free software principles in the context of generative AI.

*4.1. Legal Analysis*

Consider the case of a developer trying to prevent their publicly published work from being used to train generative AI models. What legal tools are available to this person, and under what circumstances can they be used?

Under United States (US) copyright law, authors of original works receive several exclusive rights, codified in 17 U.S.C. § 106. These include the rights to reproduce the work, prepare derivative works based on it, distribute copies to the public, and publicly perform or display the work. However, these rights are subject to statutory limitations, including exceptions such as fair use and compulsory licensing schemes that serve broader social and economic interests. Copyright automatically applies to original works when they are fixed in a tangible medium, including digital formats such as online content. This means that any original code published online is immediately protected. Licenses specify the rights that are legally reserved for the copyright holder and grant other rights to users of the work, often with attached terms and conditions. In the case of copyleft, this arrangement is reversed via a license, with all rights granted to the user on the condition that derivative works are released under the same license.

Fair use is the primary limitation of licenses serving as a method to grant developers greater control over how their code is used. Licenses, as contracts between the copyright owner and the software user, can only place terms on rights reserved for the copyright holder. Under US copyright law, the doctrine of fair use, determined by a set of four standards, allows an existing work to be used to create a new work such that it is no longer considered a derivative work of the original if the new work is sufficiently "transformative." Therefore, the original copyright holder cannot impose conditions on how their work is used if the new work qualifies as fair use.

Regarding generative AI, if it becomes established through the courts or legislation that using code to train a model is always fair use of that training data, the discussion of licenses in this context becomes irrelevant. Licenses would be unable to restrict what models can train on, given that publicly available code is unrestricted. However, there is uncertainty about whether training models on code is, or should be, regarded as fair use (Charlesworth 2024). For instance, some argue that the substantially similar relationship between the input and output of a model, the commercial purpose of many AI, models, and how models can replace the source material in the market suggest that training on copyrighted material may not qualify as fair use (Shen 2024).

The May 2025 US Copyright Office report provides crucial (although non-binding) guidance on this uncertainty. The report concludes that training AI models on copyrighted works "clearly implicates the right of reproduction," explicitly rejecting any blanket fair use protection for AI training (US Copyright Office 2025). The Office emphasized that the reproduction right is triggered when copyrighted works are copied into training datasets, regardless of whether those copies are retained after the training process is complete. This finding directly challenges the argument that AI training constitutes non-expressive use, exempt from copyright concerns. The report further notes that commercial uses creating competing content are less likely to qualify as fair use, while research and non-commercial uses receive more favorable treatment—a distinction particularly relevant for commercial code generation models.

With the assumption that this sort of training is not fair use, we can hypothesize that developers can use licenses to restrict how their code is used to train generative AI models. In this case, generative AI models trained on code could plausibly be considered derivative works of the training data, although this interpretation remains legally unsettled, as the exact code from the training data is not present in the model's codebase and is never technically executed by the model. A similar conclusion as to the ability of open source licenses to propagate to AI models has been found where Benhamou argues "when the AI model incorporates open code into its program or open data into its training dataset, there is a risk of propagation, *i.e.* a risk that the entire model or the input data become fully open, as the AI model or the input can be considered as derivative of the open code or data, whose license requires derivatives to remain open" (Benhamou 2024).

The uncertainty surrounding whether the use of open source code as training data constitutes fair use warrants consideration in its own right. The current confusion in the legal landscape naturally leads developers to restrict licenses defensively, potentially hindering both AI innovation and open source collaboration. A clear, purpose-built license can overcome this confusion by providing explicit guidance that operates effectively under multiple legal scenarios. If training is not deemed fair use, then licenses become the primary mechanism for controlling the use of AI training. A license specifically addressing AI training for code generation models would provide clear guidance to all parties, potentially increasing the pool of code available for training while respecting developer preferences. However, if courts ultimately determine that training AI models constitutes fair use, such licensing terms would have no legal force. And it is worth noting that jurisdictions outside the US, such as the EU, provide explicit text and data mining (TDM) exceptions, which permit data mining unless right-holders explicitly opt-out through licensing terms. Thus, the CCAI license terms could effectively serve as an explicit opt-out of such statutory exceptions under EU law. Therefore, we see legal value in such a license, regardless of how the US courts ultimately rule on the question of fair use.

*4.2. Policy Analysis in Each Context*

The FOSS community generally agrees on the benefits of FOSS, which, according to the community, include reliability, security, and low cost (AlMarzouq et al. 2005). However, sub-movements of the community disagree on the intellectual justification for FOSS. The free software (FS) movement presents moral claims to advocate for free software as defined above. Then, there is the open source (OS) movement, which offers pragmatic arguments regarding the value of free software. The differences can be summed up as "open source is a development methodology; free software is a social movement," but their "practical recommendations" are largely similar (Stallman 2002).

The FS movement argues for free software based on what Richard Stallman describes as the "Golden Rule" applied to software development. He argues that proprietary software is destructive insofar as it "reduces the amount and the ways that a program can be used" and, as a result, reduces the wealth humanity derives from the program (Stallman 2002). He ties this to Kantian ethics, arguing that a good citizen does not use destructive means, such as proprietary software, to become wealthier. With this, he concludes that "programmers have the duty to encourage others to share, redistribute, study, and improve the software we write" (Stallman 2002). Another way that this has been formulated is to start from the premise that software has permeated society, and, under a proprietary model, software producers can control the availability of and access to software. As a result, "we are beholden to these producers to run our modern society and are theoretically at their whim" (Heyns 2012). Proponents of FS argue that this outcome is undesirable. Instead, software should be regarded as infrastructure, as it is essential for societal participation and engagement. Thus, the assertion is made that society should regulate access to the software "infrastructure" rather than software companies. This approach helps prevent what has been termed "the digital divide," wherein select groups of people have access to the broad digital ecosystem (James 2003).

The OS movement advocates for free software from a strictly pragmatic or utilitarian framework, arguing that free software is more beneficial to society than proprietary software. Software plays a critical role today, making society interested in the quality and reliability of the software that is produced and used. OS advocates argue that making source code freely available online for public collaboration results in high-quality and highly reliable software. The OS community generally posits that this is due to the ability to peer review software and provide more timely fixes because more developers can review source code in search of bugs and security risks. When these issues are identified, a distributed community can work together to create solutions, rather than a select group within a specific company addressing the problem. The idea that a distributed community can lead to better software outcomes is summed up in the famous Linus's Law: "Given enough eyeballs, all bugs are shallow" (Raymond 1999). OS is distinct from FS in that it does not make moral claims, but is based on pragmatic claims regarding which development model will yield the highest quality and most reliable software.

The primary risks of releasing certain pieces of traditional software as open source include security vulnerabilities and software misuse (Zajdel, Costa, and Mili 2022). Regarding security vulnerabilities, the risk of malicious actors discovering and exploiting these vulnerabilities can be mitigated by allowing multiple developers to review the code and identify and fix bugs. Thus, "open source does not pose any significant barriers to security but rather reinforces sound security practices by involving many people that expose bugs quickly and offers side-effects that provide customers and the community with concrete examples of reusable, secure, and working code" (Clarke, Dorwin, and Nash 2009).

Moving from the traditional software context to generative AI, the risk profile changes. The security risks of open source AI are non-negligible, with a survey revealing that 10% of IT decision-makers have accidentally installed malicious code in their organizations through the use of open source AI tools.[5] Furthermore, the risk of actors using the code for malicious purposes is a concern that is less directly countered by the benefits of open source software and is more challenging to mitigate. Once open source software is out in the world, there is by definition no way that the developers can control its use; while Meta's Llama terms of service ostensibly prohibit its use for military purposes, researchers associated with the Chinese People's Liberation Army were still able to use it to create a tool to gather and synthesize intelligence (Wilson and Hine 2025).

Additional risks associated with open source AI include misuse, where software is used in a technically unintended manner to inflict harm. When it comes to OSS related to generative AI, there are several pathways to explicit harm, including the exploitation of security vulnerabilities and the ability to create harmful content, particularly given the challenges of rolling back or updating models (Eiras, Petrov, Vidgen, Witt, et al. 2024). The possibility of generating harmful content or using a foundation model for malicious activities is particularly concerning given the nature of such misuse.

---

[5] https://www.anaconda.com/lp/state-of-enterprise-open-source-ai

This type of misuse is primarily direct, contrasting with actor-dependent misuse seen in traditional software. For example, if a foundation model is fine-tuned on a company's emails, it may generate highly effective phishing emails, posing a more significant threat to unsuspecting employees. This scenario differs fundamentally from using an open source email client to send phishing emails. Here, the open source technology in the case of generative AI is being used to amplify harm. In contrast, in the traditional software case, the open source software operates as intended but is misused by the actor to inflict damage. The risk of this explicit misuse in generative AI is significantly higher compared to traditional software.

*4.3 Synthesizing the Analysis*

The legal analysis has shown that the CCAI license can propagate from training data to a generative AI model itself, while the policy analysis has shown strong support for the open-sourcing of traditional software, highlighting both substantial benefits and risks in the context of open source AI. The particular risk of direct misuse stands out as a significant concern, particularly in the context of open source AI, compared to traditional software, which could complicate the justification for open source approaches in the generative AI context. In this section, we will take these insights and show that the CCAI license could benefit both the open source and AI development communities when used in particular regulatory schemes. The promise such a license holds for both communities is first argued, and then a discussion of regulatory schemes follows.

This type of copyleft license offers three key advantages. First, it would give developers enhanced control over how their code is incorporated into generative AI development. Second, it would promote the development of genuinely open source generative AI systems. Third, it would help combat "open-washing" practices that dilute the real benefits that open source generative AI systems can provide.

First, a copyleft license that propagates from traditional software to generative AI systems provides an important mechanism to give developers greater control over how their code is used in AI systems. Today, when developers open source their code, they lose control over what types of generative AI models are trained on their code and for what purposes those AI models are employed. On the one hand, this is a feature, not a bug. Open source licenses are explicitly designed to allow the unrestricted use of code, enabling others to use the original work in new and innovative ways. At the same time, developers often do not want their work to contribute to nefarious ends, which could occur if their code is used to train a generative AI model intended to cause harm, for example. This has led some developers to add "no AI" clauses to existing licenses to prohibit using their code for training an AI model.[6]

The tension within the open source community regarding the balance between freely sharing software and caution towards unrestricted training on such software is evident in the backlash against the release of CoPilot. GitHub CoPilot is an AI model developed specifically to generate code and, consequently, requires a vast amount of code for its training data. This prompted GitHub to use code from numerous open source repositories, asserting that the training of their model constituted fair use and, therefore, the licensing terms of the code did not apply. Many in the FOSS community disagreed, leading to a boycott of GitHub by many open source developers and a class-action lawsuit against GitHub, Microsoft, OpenAI, and others involved in creating CoPilot (Class Action Complaint 3:22-cv-06823). Some even argue that foundational models like CoPilot may discourage developers from contributing to open source projects, as these models "exploit the principle of solidarity operating within the movement" and have the potential to undermine the movement altogether (Novobilská 2023). An open source license with a copyleft clause would specify that it propagates from training data to generative AI models, providing open source developers a middle ground between allowing unconstrained training and limiting innovation and openness by restricting all training.

Second, introducing and adopting a copyleft license that propagates from training data to AI models would lead to more open source generative AI models—an outcome that the open source community supports.[7] It is well known that training generative AI models requires vast amounts of data, so a model designed to generate code (such as GitHub's CoPilot) would need substantial code to train on. Open source repositories are currently the solution for model developers. Suppose such a license were widely adopted within the FOSS community. Of course, the effectiveness of the CCAI license depends heavily on its widespread adoption. If only a small subset of developers adopts it, proprietary AI developers may simply avoid CCAI-licensed code. A license of this kind would outline which portion of the open source code online is available for use by open source models during training. This would create an incentive for the open sourcing of new generative AI models that produce code, as it provides open source model developers with access to training data that proprietary models cannot use. The incentivization and creation of open source generative AI models offer new opportunities for external oversight, technical innovation, and decentralized control over AI development compared to closed-source models (Seger et al. 2023).

---

[6] https://github.com/non-ai-licenses/non-ai-licenses
[7] https://opensource.org/ai/open-source-ai-definition

Third, a copyleft license of this type could provide more explicit, enforceable definitions for open source AI, helping to combat "open-washing." Open-washing refers to AI models that claim to be open source by publicly sharing only one component, such as the model weights (Liesenfeld and Dingemanse 2024). While this may allow users to download and run the model on their computers, it fails to convey the comprehensive understanding of software freedom that the open source community has cultivated; crucial contexts, including the code that trained the model and details of the training data, are absent. The current language in copyleft licenses mandates that the "entire work" be shared under the same license as the original, not just the portion of the software copied from the original work. From a model developer's perspective, the "entire work" could encompass not only the model weights but also the underlying model architecture, supporting code, and details about the training data. The interpretation of "entire work" as it applies to generative AI could broaden the required components of an AI system that must be open to tackle the open-washing observed in the AI development community effectively.

These three advantages of the CCAI license must be considered in light of the above legal and policy analysis. Specifically, the risk of direct misuse of open source AI is that it is used for malicious purposes. The challenge is that open source licenses must not restrict how a technology can be used, as defined by the principles of open source.[8] However, while it can't be prohibited, there is precedent within the FOSS community resisting the misuse of open source software. For instance, a former developer at Chef, a computer management software company, created open source tools that helped customers work with Chef's software. After leaving the company, he discovered a contract between Chef and the US Immigration and Customs Enforcement (ICE), prompting him to remove the code with a note stating, "I have a moral and ethical obligation to prevent my source [code] from being used for evil."[9] This decision caused an outage for multiple customers, leading Chef to republish an older version of the open source tool's source code on the company's corporate GitHub account, an action permitted by the software freedom provisions of the original license. However, it should be noted that the developer was able to protest Chef's collaboration with ICE because he had authority over that code.

One possible way to mitigate the risks of open source generative AI would be to introduce a responsible AI use regulation that limits how open source models can be used to cause direct harm. For instance, part of the EU AI Act prevents AI systems from using subliminal, manipulative, or deceptive techniques that distort a person's behavior and impair decision-making, causing significant harm, or exploiting individuals or groups based on age, disability, or socio-economic conditions to influence behaviors in a harmful way (European Union 2024). Although these provisions have been criticized for a lack of specificity (Montag and Finck 2024), similar restrictions in the US could help mitigate the risk of misuse of open source generative AI systems. This would help rebalance the risks and benefits of open source generative AI closer to those of traditional open source software, allowing for a stronger endorsement of copyleft in both contexts.

---

[8] https://opensource.org/osd
[9] https://www.zdnet.com/article/no-source-code-for-evil-developers-pressure-chef-software-to-cut-ties-with-ice/

## 5. Conclusion

The analysis presented in this article demonstrates that the Contextual Copyleft AI (CCAI) license represents a promising approach to extending traditional copyleft principles into the domain of generative AI development. Through our structured evaluation framework, we have determined that such a license is legally feasible under current copyright law, assuming that AI training does not constitute fair use, and offers substantial benefits to both the open source and AI development communities. The license addresses three critical challenges facing the FOSS community in the age of generative AI: providing developers with meaningful control over how their code is incorporated into AI systems, creating incentives for genuinely open source AI development, and combating the dilution of open source principles through open-washing practices. These advantages align with the fundamental goals of the FOSS movement while adapting to the unique characteristics of generative AI technologies.

However, our policy analysis reveals that the risk profile of open source generative AI differs significantly from that of traditional software, particularly regarding the potential for direct misuse, where AI systems can be employed to generate harmful content or amplify malicious activities. This increased risk profile challenges the traditional cost-benefit calculus that has historically justified copyleft licensing in software development. The analysis suggests that while the benefits of open source AI remain substantial—including enhanced oversight, innovation, and decentralized control— the heightened risks require careful consideration and potentially complementary regulatory mechanisms to achieve an appropriate balance.

The path forward for implementing the CCAI license depends critically on the development of comprehensive regulatory frameworks that address the responsible use of AI systems. Such regulations, exemplified by provisions in the EU AI Act that restrict manipulative and exploitative AI applications, could help mitigate the direct misuse risks that complicate the justification for open source AI. CCAI alone does not eliminate misuse risks—it ensures openness but cannot prevent malicious use of openly released models. Thus, its practical effectiveness depends significantly on complementary regulatory measures addressing responsible use of AI. Within such a regulatory environment, the CCAI license emerges as a valuable tool for preserving the core values of software freedom while adapting to the realities of modern AI development. This analysis thus contributes to the broader discourse on maintaining the integrity and relevance of FOSS principles as technology continues to evolve, demonstrating that thoughtful adaptation of traditional licensing mechanisms can help navigate the complex challenges posed by generative AI while upholding the fundamental commitments of the open source community.

**Appendix A: The CCAI License**

The following is the text of the CCAI license designed to propagate from training data to generative AI models.

>  Contextual Copyleft AI (CCAI) License v1.0
>
> Copyright (c) [year] [author]
>
> Permission is hereby granted, free of charge, to any person obtaining a copy of this software and associated documentation files (the "Software"), to use, copy, modify, merge, publish, distribute, and/or sublicense copies of the Software, subject to the following conditions:
>
> *1. Software Freedom.*
> Users are granted the freedom to run, study, modify, and distribute the Software, including modified versions, for any purpose.
>
> *2. Copyleft Requirement.*
> Any distribution, publication, or making available of the Software — including exact copies, modified versions, or derivative works — must be licensed in full under the terms of this CCAI License without imposing any further restrictions. This requirement also applies to any model, dataset, or system that is trained on, derived from, or incorporates the Software or its outputs. In the case of generative AI models, the obligation to apply this license includes, at minimum, the disclosure of the model's source code, a meaningful description of the training data, and the model's parameters, weights, and architecture.
>
> *3. Distribution Clarification.*
> Providing access to the Software or any covered model over a network or online service shall be considered distribution and triggers the obligations of this License.
>
> *4. No Warranty.*
> The Software is provided "as is", without warranty of any kind.
>
> *5. Preservation of License Notice.*
> Copies, modifications, and derivative works must retain this license notice.

**Appendix B: The CCAI Clause**

The following is the text of the CCAI clause that can be added to existing open source licenses to propagate the license from training data to generative AI models.

> Contextual Copyleft AI (CCAI) Clause
>
> If this software is used to train, fine-tune, or otherwise generate any machine learning model or generative AI system, any resulting model, dataset, or system must be released under terms no more restrictive than this license. This includes the requirement to make available the training code, a description of the data used in training, and the model's parameters and architecture. Providing access to such a model or its outputs via a network is considered distribution and triggers this obligation.